\documentclass[11pt, conference]{IEEEtran}

\usepackage[pdftex]{graphicx}
\usepackage[cmex10]{amsmath}
\usepackage{array}
\usepackage{amsmath}
\usepackage{amsfonts}
\usepackage{amssymb}
\usepackage{graphicx}%
\usepackage[bottom]{footmisc}

\providecommand{\U}[1]{\protect\rule{.1in}{.1in}}

\begin{document}
	 
	\title{O.D.E.S. : An Online Dynamic Examination System based on a CMS Wordpress plugin}

	\author
	{
		\IEEEauthorblockN{George F. Fragulis\IEEEauthorrefmark{1},
			Lazaros Lazaridis\IEEEauthorrefmark{1},
			\\Maria Papatsimouli\IEEEauthorrefmark{1}    
			and
			Ioannis A. Skordas\IEEEauthorrefmark{1}}\\

		\IEEEauthorblockA{\IEEEauthorrefmark{1}Laboratory of Web Technologies \& Applied Control Systems\\ Dept. Of Electrical Engineering\\
			Western Macedonia Univ. of Applied Sciences, Kozani, Hellas 
			}

	}

	\maketitle
	
	\begin{abstract}
		 
		 This paper describes the online dynamic examination application plugin named "ODES", developed according to the open source software philosophy, where the CMS Wordpress is used as programmers/coders are given the potential to develop applications from scratch with safety and ease.
		 In ODES application there exists two types of users: admin/teacher and student. The admin/teacher can create/edit/delete/view questions, categories of questions and examination papers. 
		 The questions are divided in two types, multiple choice questions (including true/false) and long answer questions (essays).   The teacher can create exams choosing the number of questions and the types of questions that exist in the pool of questions that are previously created. The selection is done randomly by the application and the teacher just determines the total number of both multiple choice or long answer questions as well as the importance (weight) of each one of them (including negative grades also). The student takes the random generated exam and receives his/hers grades. The grades of the multiple choice questions are done automatically, whereas for the long answer questions the teacher  is responsible to put grades on. After the completion of the exam the teacher can view the student's final score via the control panel or a report.
		 

	\end{abstract}
	
	\begin{IEEEkeywords}
		Online examination system; Content Management; Wordpress plugin; Web Based Application; intelligent system; automated transaction 
	\end{IEEEkeywords}

	%
	\IEEEpeerreviewmaketitle

\section{Introduction}

In the recent years,  as a result of the development of Internet technology, online examination systems have gained remarkable  preference and acceptance in the education industry because  they have modernized the process in which students prepare 
and appear for their examinations \cite{younis2015construction}.  Also there has been a growing interest of light open source content management systems (CMS) such as WordPress or Drupal, that configured in an appropriate way, allow a teacher to set up quickly and easily an open environment for a virtual classroom/examination site. 

\subsection{Content Management Systems}

A content management system (CMS) is a software that help us make websites easier, faster, browser compatible  and responsive  websites in  a  very  short  time with  the  powerful inbuilt  features. The basic advantages of the most popular CMS are the following:

\begin{itemize}
\item Security
\item Usability
\item Documentation / Support
\item Expansion and correction
\item Easy administration / installation
\end{itemize}

\subsection{Wordpress}

Wordpress is a free and open source web software tool and a content management system (CMS) based on PHP and MySQL. It has  many  features  including  a  plugin  architecture  and  a  template  system. Wordpress is  among  the  most  popular systems  in  use  on  the  Internet. 
Wordpress provides security and easy-to-use programming software both for developers and website administrators  and there are thousands of developers who contribute to its source code optimization \cite{Fernandes2015}.
Plugins are tools which extend the functionality of Wordpress. The core of WordPress is designed to be clean, to maximize flexibility and minimize code bloat. Plugins offer custom functions and features so that each user can tailor their site to their specific needs \cite{Koskinen2012}, \cite{Williams2011}, \cite{hills2016navigating}. The functions that provided by Wordpress are used in order the development to be done in its own environment. The application form is the plugin so  both the installation/uninstallation can be done easily by the users.

\subsection{Literature review}
In this section we examine the related works on the online examination systems.
In education, the concept of E-Learning  has grown rapidly from distance 
learning to virtual classrooms towards the online courses and online examinations. Universities and Colleges  are trying to move from a paper-based to a paperless environment \cite{younis2015construction}, \cite{Rosenberg2002}, 
\cite{Driscoll2002}, \cite{Welsh2003}, \cite{Garrison2011}, \cite{Sun2008}, \cite{Zhang2004}, \cite{Tavangarian2004}, \cite{Hrastinski2008}, \cite{Levy2017}, \cite{Al-Samarraie2017},\cite{Shen2006},\cite{Trivedi2010}.

Many researchers proposed a number of online examination systems such as SIETTE \cite{Guzman2005}, WETAS \cite{Henke2006}, EMS \cite{Rashad2010},WONES \cite{Sheshadri2012} and CBTS \cite{Fagbola2013}. A number of application results using other open source programming languages such as PHP and MySQL can be found in  \cite{Skordas2011}, \cite{Skordas2014},\cite{Lazaridis2016},\cite{Skordas2017}.

\section{Application description}

Our proposed plugin application named (ODES) is developed for the Wordpress platform.
Through this specific plugin the roles of the teacher and the student are
created. The teacher inserts either multiple choice or long answer
questions (essays) and can create exams. In each exam the teacher defines only the
importance (weight/score) that each question will have (including negative values if needed). 

The exam generation is based on a mixing of long answer and multiple choice questions which are selected in order to be displayed according the specification that entered by the teacher .  
\begin{figure}[b]
	\centering
	\includegraphics[scale=0.3]{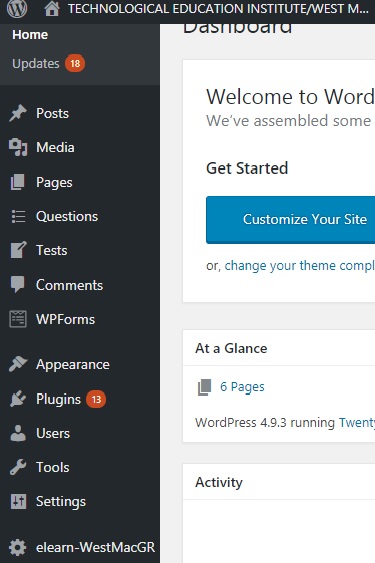} \caption{Application options have
		been added in the menu}%
	\label{fig:eik2}%
\end{figure}
The selection of the exam questions is done randomly.
In  the case of multiple choice questions, both questions and answers of each
one are randomly displayed for different users. Multiple choice questions are automatically corrected by the system. On the other hand, there are long answer questions which are also displayed randomly for different users but the final score is given by the teacher himself.

By the time the exam is completed, the teacher is
asked to check the long answer questions (if any) so as to give the grades for
them. The final exam score(grades) appears automatically in the system for all students that take the exam with time stamp for each one of them.

\subsection{Plugin installation and activation}

Like most Wordpress functions, the plugin installation does not
require any special knowledge on programming as the only thing that is required is
the connection to the control panel and then the plugin is inserted (Plugins
-%
$>$
Add new). In this case all the available plugins can be viewed and the
selected one is loaded. The downloading procedure can be done
with the use of any file transfer program. The uploading of
plugin can take place in a zip form in the plugin's folder
(/wp-content/plugins/) and it will be loaded automatically by
Wordpress. Finally, the activation must be done manually.

Initially in the administration menu (Figure \ref{fig:eik2}), there are three options  the
\textquotedblleft Questions\textquotedblright, \textquotedblleft
Tests\textquotedblright\ and \textquotedblleft
elearn-WestMacGR\textquotedblright.

\begin{figure}[t]
	\centering
	\includegraphics[scale=0.5]{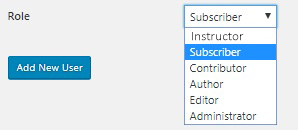} \caption{Adding the Instructor user entity}%
	\label{fig:eik4}%
\end{figure}

- A new user entity can be created with the name \textquotedblleft
instructor\textquotedblright\ (Figure \ref{fig:eik4}) that has administrator rights.

\begin{figure}[b]
\centering
\includegraphics[scale=0.27]{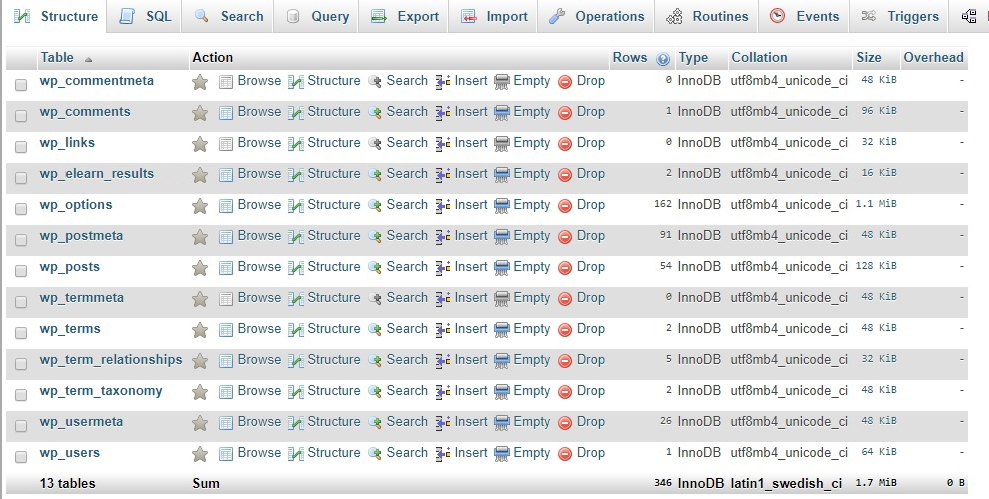} \caption{The table
wp\_elearn\_results of the database}%
\label{fig:eik3}%
\end{figure}

- In the database, a new table is added with the name
wp\_elearn\_results in which the students' answers will be stored (see Figure \ref{fig:eik3}).

\subsection{Application settings}

\begin{figure}[t]
\centering
\includegraphics[scale=0.3]{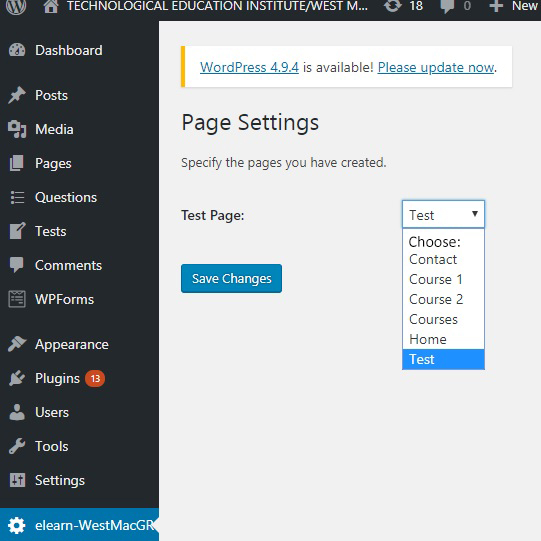} \caption{Application settings}%
\label{fig:eik5}%
\end{figure}

The application enables us to select in which page of our website we want all
the exams to be displayed. From the popup menu we can select one of the pages
that exist in the Wordpress and is defined as exam page (Figure \ref{fig:eik5}).

The page settings can be found in file settings.php via Wordpress function add\_menu\_page

\subsection{Questions}

Questions are an custom post type which is defined in
cpt-erotisi.php file by making use of register\_post\_type Wordpress
function. Even though questions are posted they cannot be seen by anybody
because during their definition the \textit{public setting} is turned to false. They are only available through the control panel.

Every question belongs to one or more question categories that are also
defined by cpt-erotisi.php file and it is a classification that is defined
through register\_taxonomy Wordpress function.

\begin{figure}[h]
\centering
\includegraphics[scale=0.23]{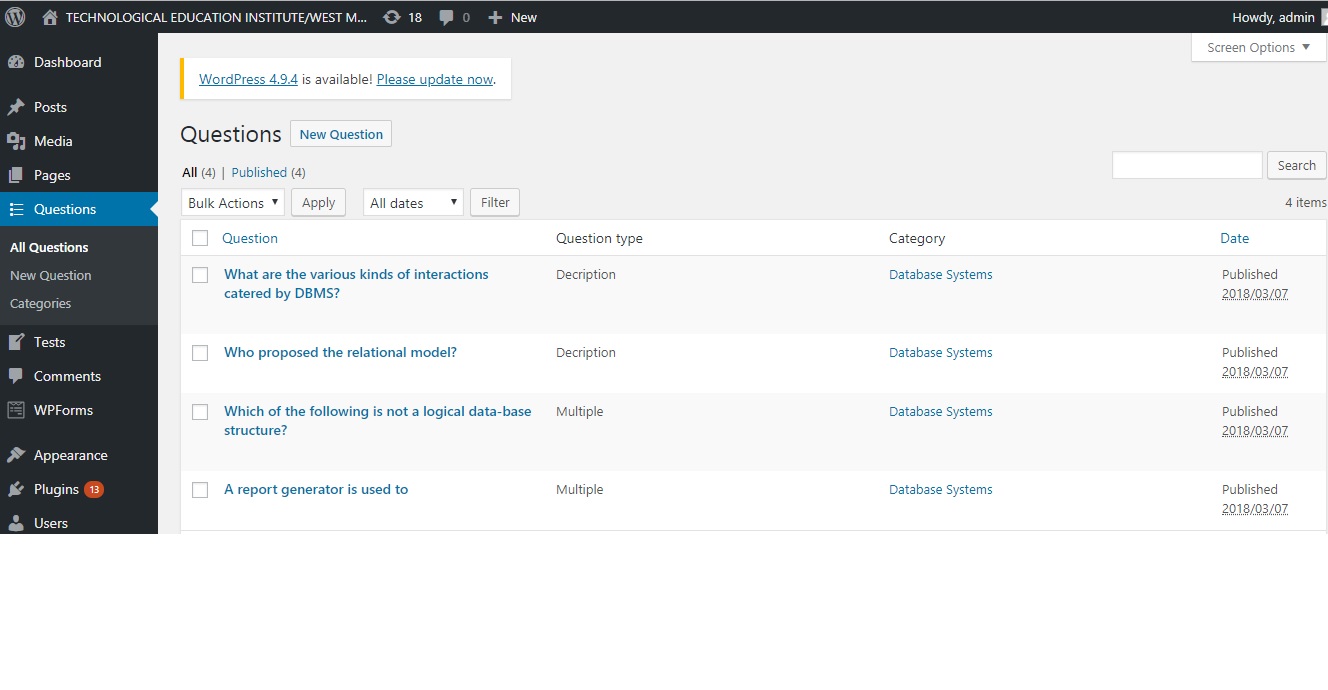} \caption{Custom post type
questions}%
\label{fig:eik6}%
\end{figure}

By pressing the menu \textquotedblleft Questions\textquotedblright\ all the
stored questions are displayed in a table where the \textquotedblleft
Title\textquotedblright, \textquotedblleft Type\textquotedblright,
\textquotedblleft Category\textquotedblright\ and \textquotedblleft entry
date\textquotedblright\ of each question are viewed (Figure \ref{fig:eik6}).

\subsubsection{Question adding}

\begin{figure}[t]
\centering
\includegraphics[scale=0.27]{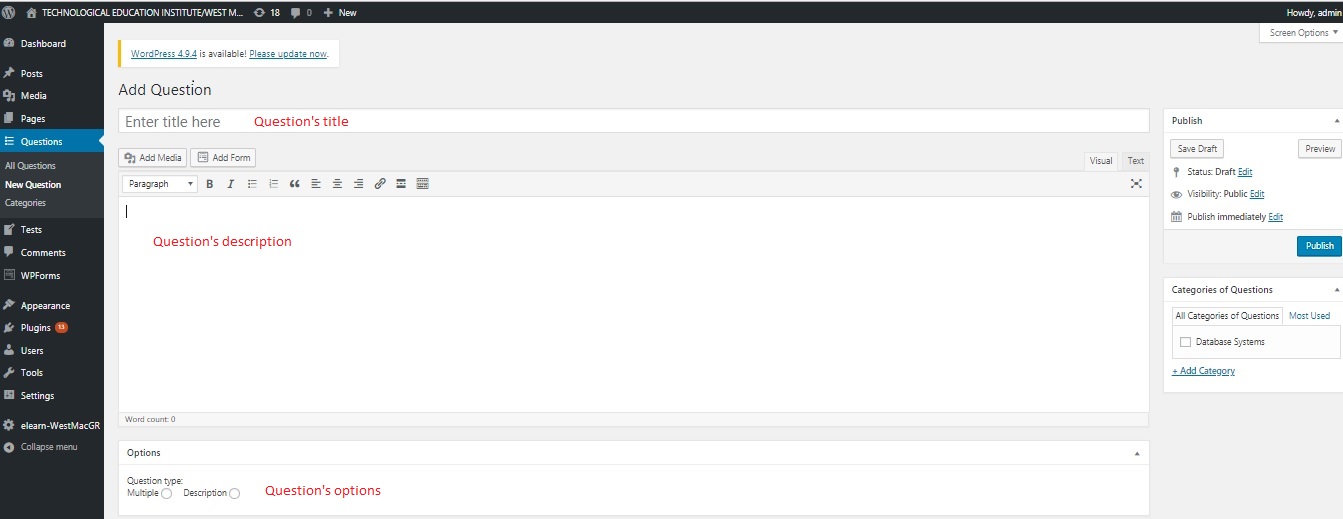} \caption{New question adding}%
\label{fig:eik7}%
\end{figure}

In order to add a new question we need to select the submenu
\textquotedblleft New Question\textquotedblright . Then a form will be
displayed for completion (Figure \ref{fig:eik7}). This form is almost the same with the standard form
that exists in the Wordpress in order to post an article or a page.

In the form there are the following fields : 
\begin{itemize}
	\item Title
\item Question description
\item Question options
\item Categories
\item Publication
\end{itemize}

In the Title field we complete the main question that the student must answer.

In the Description field the general instructions of the question are
optional. This field can accept formatted text as well as multimedia files (images,
video, etc.). 

\begin{figure}[h]
	\centering
	\includegraphics[scale=0.3]{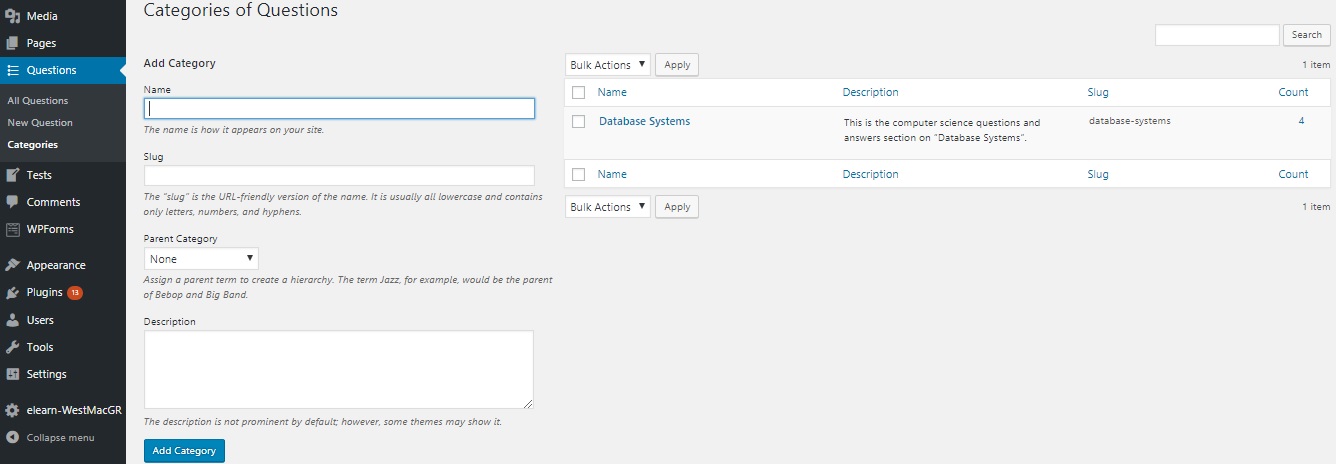} \caption{Question category adding}%
	\label{fig:eik9}%
\end{figure}

In the Categories field (Figure \ref{fig:eik9}) we can select in which of the existed categories or 

\begin{figure}[b]
	\centering
	\includegraphics[scale=0.25]{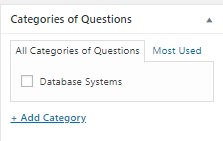} \caption{Quick question category
		add}%
	\label{fig:eik8}%
\end{figure}

sub-categories the question belongs. In the case we want to add a new category we must press the link \textquotedblleft+
Add category\textquotedblright. In the first field we enter the title of the category, in the popup menu -- parent category -- we
select if the category will be a sub-category or not and we press the button
\textquotedblleft Add category\textquotedblright as in Figure \ref{fig:eik8}.

In the \textquotedblleft Publish\textquotedblright\ field we can publish the question or
store it in order to change its settings later.

Finally, in the \textquotedblleft Options\textquotedblright\ field we define
several question elements that unlock depending on the type of question that
we want to add.

\subsubsection{Question options}
\begin{figure}[t]
	\centering
	\includegraphics[scale=0.3]{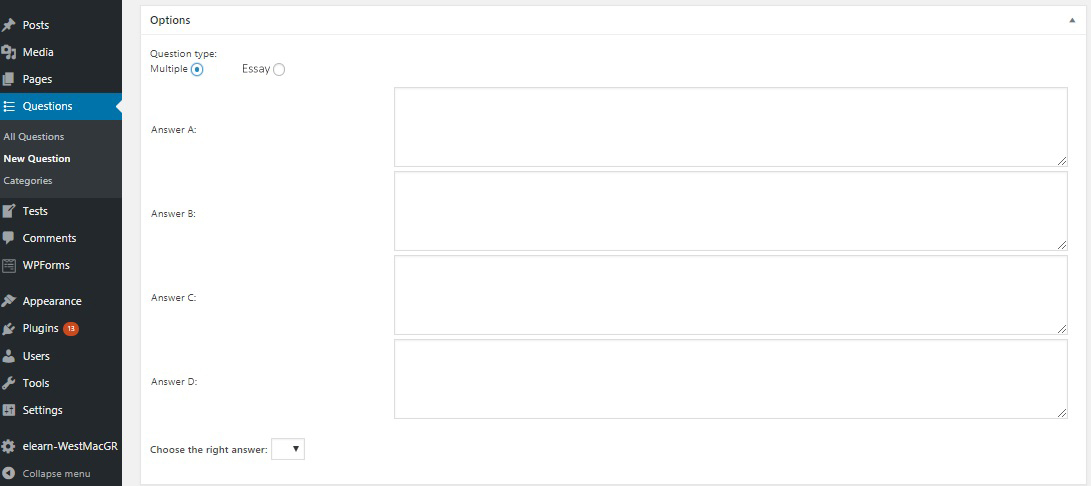} \caption{Multiple choice
		questions options}%
	\label{fig:eik10}%
\end{figure}
Initially in the \textquotedblleft Option\textquotedblright\ field we' re
required to select if the question is multiple choice or long answer. If we
select \textquotedblleft Description\textquotedblright\ then some necessary
extra options are displayed automatically. The swapping is done via jQuery
code dynamically.

As shown in Figure \ref{fig:eik10}, if we select multiple choice questions in the
\textquotedblleft Question type\textquotedblright\ we have to define the
four possible answers and then which one is the correct. Finally, we can
store or publish the question.

\subsubsection{Possible Errors during publication}

\begin{figure}[h]
\centering
\includegraphics[scale=0.27]{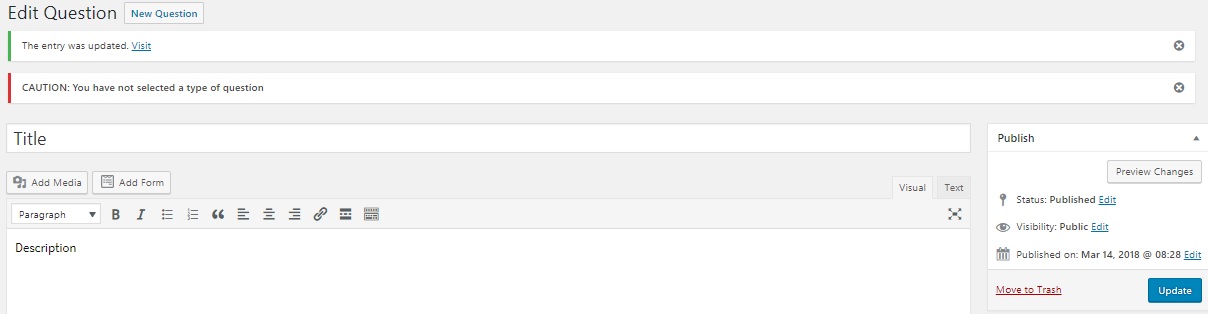} \caption{Warning for not choosing question type}%
\label{fig:eik11}%
\end{figure}

In case the question publication takes place without having chosen its type,
then the system warns us as shown in Figure \ref{fig:eik11}.

\subsubsection{Question categories}

\begin{figure}[ptb]
	\centering
	\includegraphics[scale=0.3]{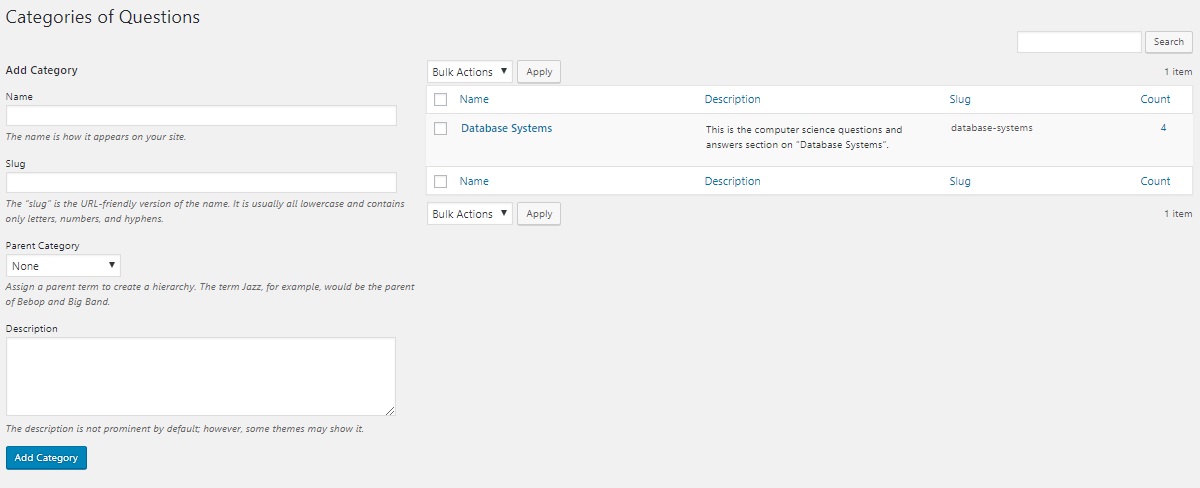} \caption{Management question
		category page}%
	\label{fig:eik12}%
\end{figure}

\begin{figure}[h]
\centering
\includegraphics[scale=0.3]{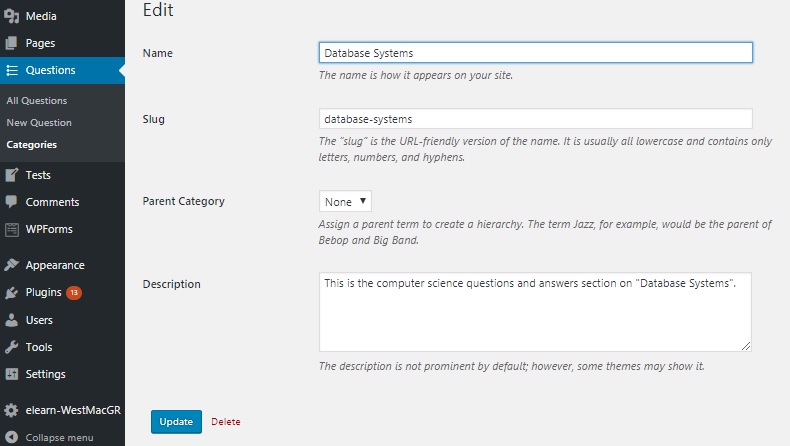} \caption{Question category
editing}%
\label{fig:eik13}%
\end{figure}

In the third sub-menu of the menu \textquotedblleft
Questions\textquotedblright, there is the question category management page (Figure \ref{fig:eik12}).
We can see all added categories and from the left page side we can add new
ones. Also, we can edit each category separately (Figure \ref{fig:eik13}).

\subsection{Exams}

Exams are, like questions, an adapted post type and are defined in the
cpt-diagonisma.php file. They are accessible to administrator/teacher user group in contrast with the questions and have their own properties.

\begin{figure}[ptb]
\centering
\includegraphics[scale=0.3]{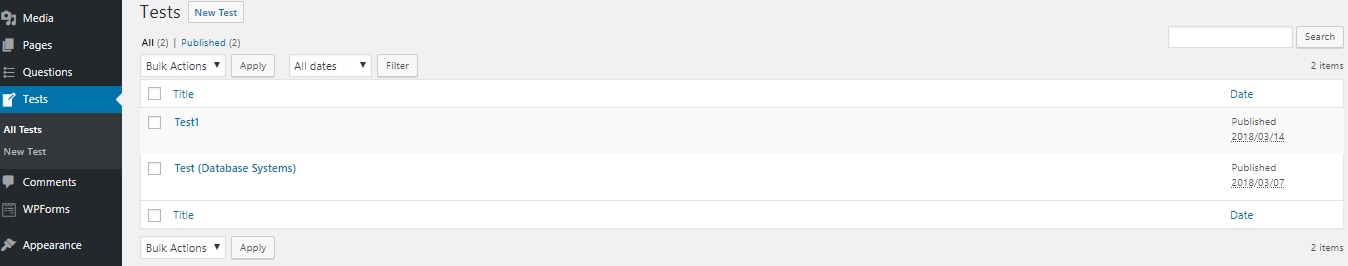} \caption{Custom post type exam}%
\label{fig:eik14}%
\end{figure}

By pressing the menu \textquotedblleft Tests\textquotedblright\ all the stored
exams are displayed in a table where the Title and Entry date are viewed.
Furthermore, the sub-menus \textquotedblleft All Tests\textquotedblright,
\textquotedblleft New Test\textquotedblright\ are viewed as well (Figure \ref{fig:eik14}). 

\bigskip

\subsubsection{Adding Exams}

\begin{figure}[t]
\centering
\includegraphics[scale=0.24]{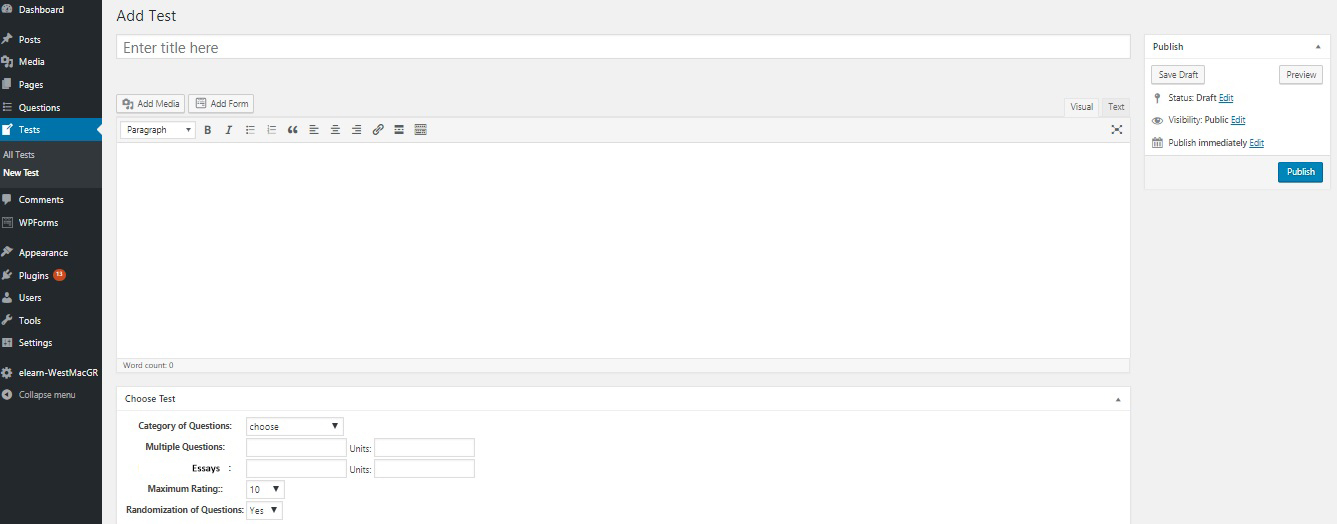} \caption{New exam addition}%
\label{fig:eik15}%
\end{figure}

In order to add a new exam we just need to press the sub-menu
\textquotedblleft New test\textquotedblright (Figure \ref{fig:eik15}).

In the new exam page the following fields can be completed:
\begin{itemize}
	\item Title
	\item Permanent link (it is displayed after the draft save)
	\item Description
	\item Options
	\item Publish
\end{itemize}

In the Title field we complete the title that we want the exam to have

In the field Permanent link we can define which the unique URL is going to be.
The options are also depended on the Wordpress permanent link settings.

In the Description field we can write general instructions concerning the
exam. This field can accept formatted text as well as multimedia (images,
video, etc.). This field can be blank.

In the Publish field we can publish or save the exam as a draft and edit it
later without being published yet.

Finally, in the Options field there are options concerning how the exam is
going to manage the questions.

\subsubsection{Exam options}

\begin{figure}[h]
	\centering
	\includegraphics[scale=0.3]{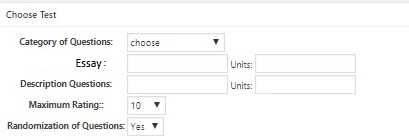} \caption{Exam options}%
	\label{fig:eik16}%
\end{figure}

Initially in the exam options the teacher  defines from which question
category the selection will take place. In the popup menu all the existing
categories are displayed automatically. Then we complete how many multiple
choice and long answer questions will be selected. Next to each selection
there are the corresponded question importance fields (Figure \ref{fig:eik16}).
The importance is about the question type and there is an equivalent
allocation on each one of them. Then, reduction to the unit by
the system is done and the scoring will be displayed with a maximum of 10 or 100
depending on what we have chosen on the popup menu \textquotedblleft Maximum
Rating\textquotedblright. Finally, we select if there will be a
question randomization. If \textquotedblleft No\textquotedblright\ is selected
then the system will display the same questions in the same order to every student.


\subsubsection{Running the Student exam}
 
As shown in the following question, depending on the options that were
previously made we can see all the available exams. The exam page is defined
in the templates/page-diagonismata.php file.

\begin{figure}[t]
\centering
\includegraphics[scale=0.3]{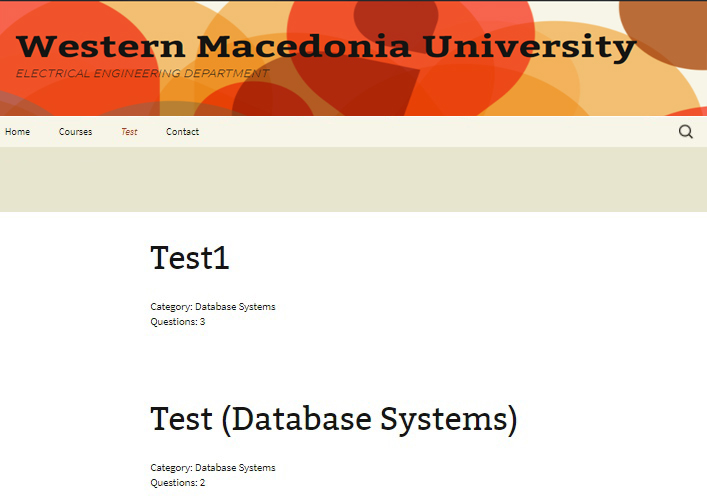} \caption{Exam page}%
\label{fig:eik17}%
\end{figure}

By pressing on one of the exams (Figure \ref{fig:eik17}) the student is redirected to his/her their exam page which is defined by tempates/single-diagonisma.php file.

Initially the student is asked to provide his personal details. But the system
has already selected the questions which will be displayed in him and will
store them along with his personal details. This happens because we don't want 
the student to view the selected questions and to prevent him from
making any unwanted actions (e.g. to close the browser tab and revisit the
exam hoping for \textquotedblleft easier\textquotedblright\ questions).
\begin{figure}[h]
	\centering
	\includegraphics[scale=0.3]{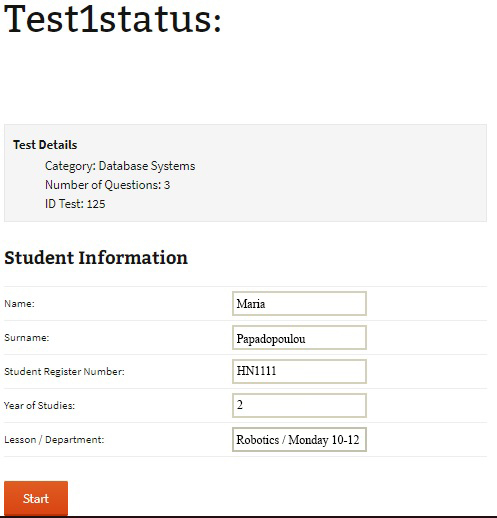} \caption{Registering student
		details for the exam}%
	\label{fig:eik18}%
\end{figure}

When the user presses the exam \textquotedblleft Start\textquotedblright\ button his/her personal details are stored in the database and
specifically in the the wp\_elearn\_results table (Figure \ref{fig:eik18}). The entry time, exam status and selected questions are also stored in a  array  in
order to be stored in a table field. Furthermore, the HTML  symbol
\textquotedblleft\ \textquotedblright' has been override by putting the
\textquotedblleft%
$\backslash$%
\textquotedblright\ symbol. This technique prevents any unwanted malicious action.

\begin{figure}[h]
	\centering
	\includegraphics[scale=0.20]{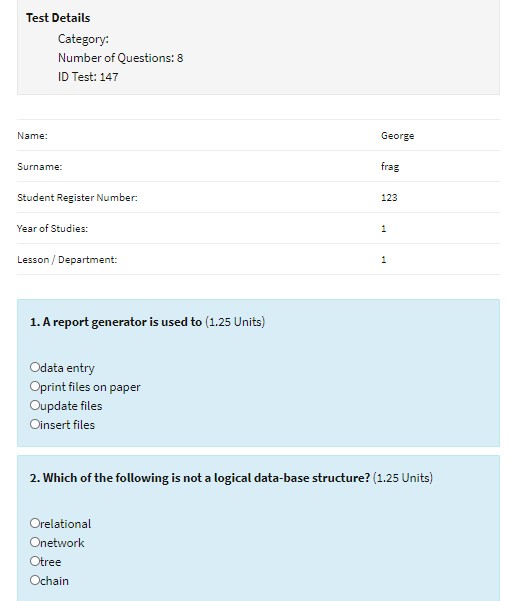} \caption{Multiple choice questions}%
	\label{fig:eikonagff1}%
\end{figure}

\begin{figure}[ptb]
	\centering
	\includegraphics[scale=0.20]{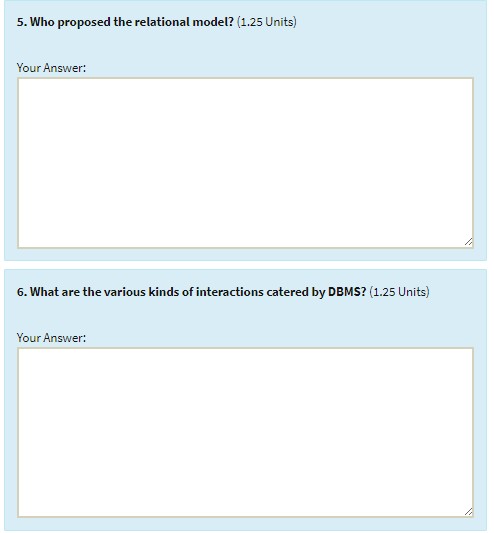} \caption{Long answer/essay form question}%
	\label{fig:eikonagff2}%
\end{figure}
Then the student  is redirected in a new web page that has all
the questions that have been selected randomly and is asked to answer them (Figures \ref{fig:eikonagff1}-\ref{fig:eikonagff2}). In
the case of multiple choice questions the possible answer order is random as
well .
\begin{figure}[ptb]
	\centering
	\includegraphics[scale=0.3]{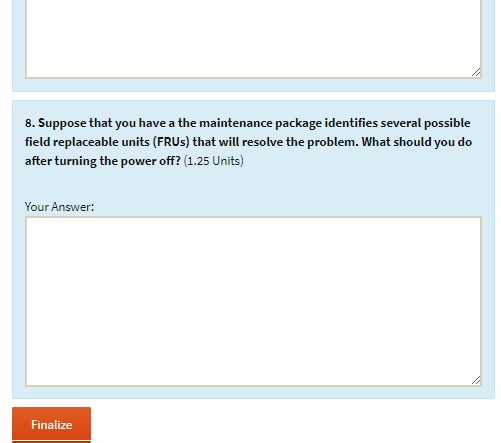} \caption{Finalize button}%
	\label{fig:eik20-21}%
\end{figure}

The student, after reading and answer the questions must press the
\textquotedblleft Finalize\textquotedblright\ button in order the exam to be stored in
the database and then the teacher will be able to check it (Figure \ref{fig:eik20-21}).

\begin{figure}[ptb]
\centering
\includegraphics[scale=0.27]{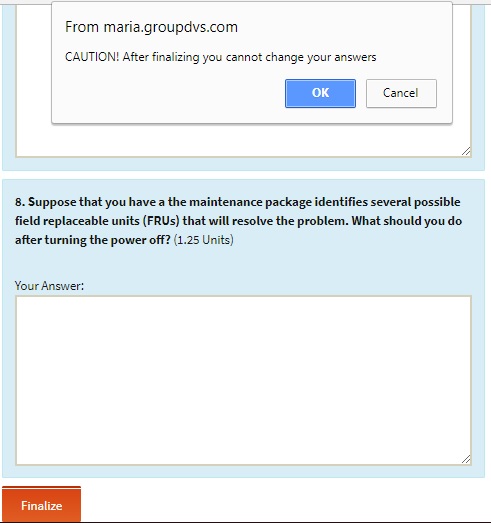} \caption{Warning message
before test submmited}%
\label{fig:eik20-21b}%
\end{figure}

Before the final answers is sent, there is a final validation message (Figure \ref{fig:eik20-21b}) warning
the student that he/she cannot be able to change the answers if the exam is submitted.

\begin{figure}[ptb]
\centering
\includegraphics[scale=0.27]{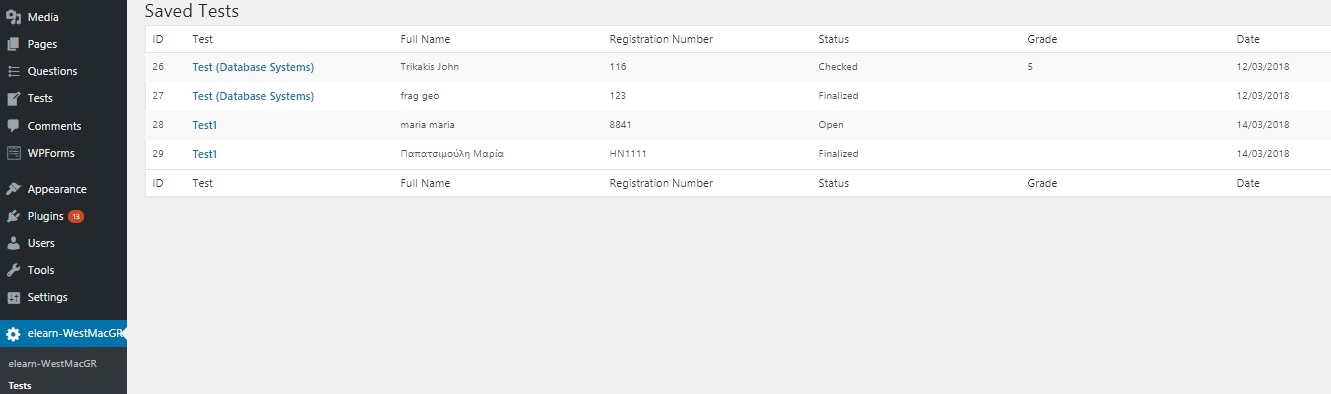} \caption{Status change from
open to finalized}%
\label{fig:eik20-21c}%
\end{figure}

The answers have been stored in the database and also the status has been
changed from \textquotedblleft open\textquotedblright\ to \textquotedblleft
finalized\textquotedblright (Figure \ref{fig:eik20-21c}).

\subsection{Answer correction by the teacher}

The teacher can view the stored answers from the menu \textquotedblleft
elearn-WestMacGR\textquotedblright\ $\rightarrow$ \textquotedblleft
Tests\textquotedblright. In the templates/admin-results.php file, there is a
table with all the answered exams. In the table the exam title, students'
personal details, exam status, grades and entry date are shown. By pressing
the exam title, the teacher can correct and put grades on.

\begin{figure}[b]
	\centering
	\includegraphics[scale=0.22]{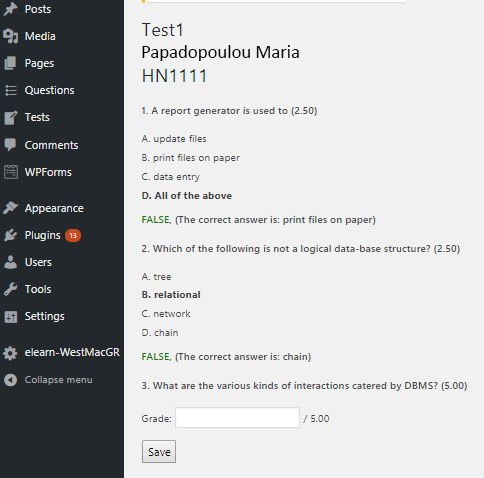} \caption{Exam check and score finalization}%
	\label{fig:eik24}%
\end{figure}

As shown in Figure \ref{fig:eik24}, in the templates/admin-results.php file the
submitted answers have already been corrected and the teacher is asked to
check the long answer questions.

\begin{figure}[b]
\centering
\includegraphics[scale=0.22]{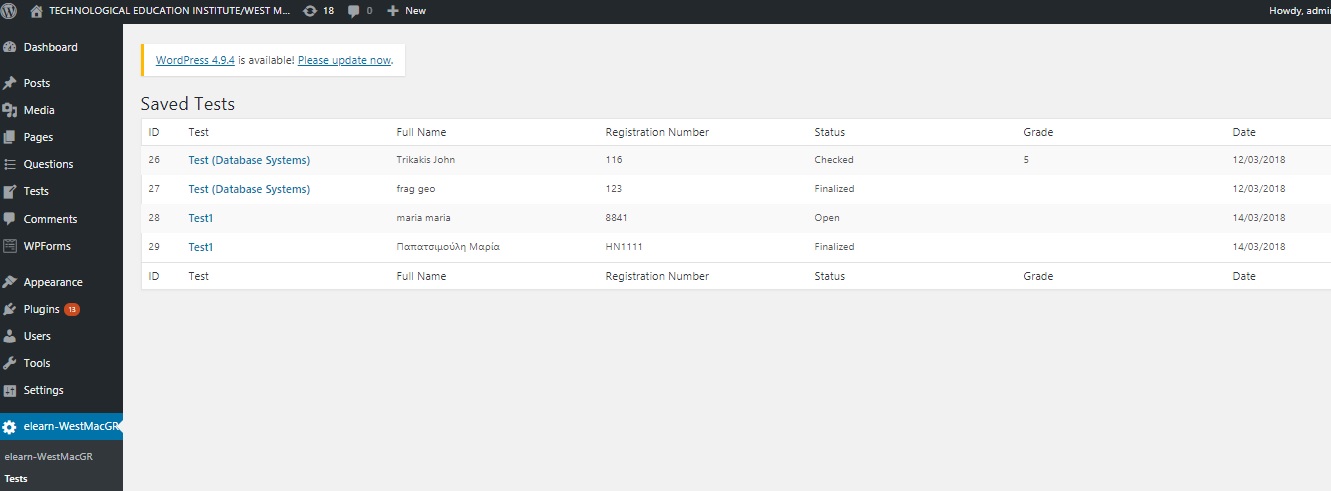} \caption{Final Grades display}%
\label{fig:eik25}%
\end{figure}

After the teacher inserts the grade in the long answer questions in the
list with the stored exams the grade is displayed for the specific exam and the
status changes to \textquotedblleft Checked\textquotedblright (Figure \ref{fig:eik25}).

\begin{figure}[h]
	\centering
	\includegraphics[scale=0.22]{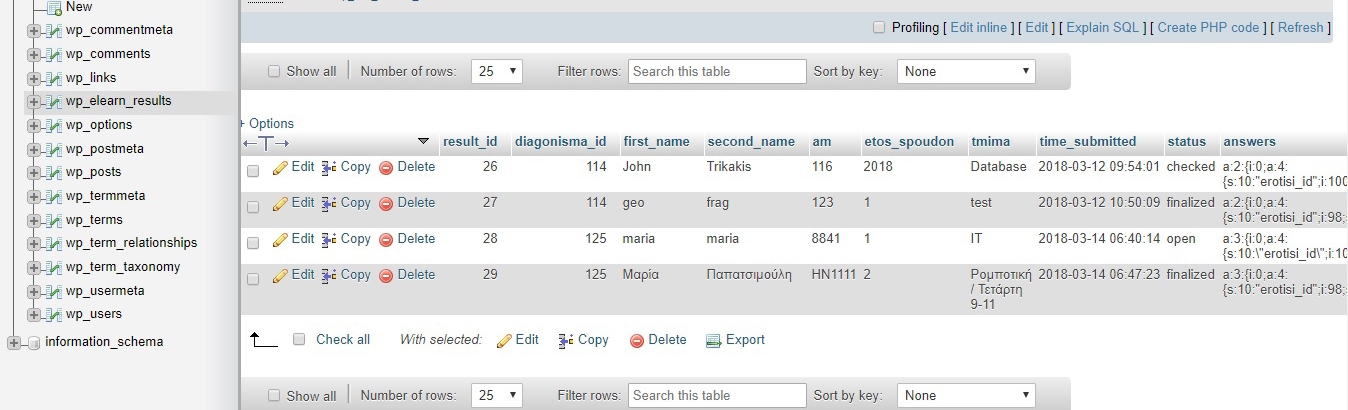} \caption{Stored exams management}%
	\label{fig:eik22}%
\end{figure}

In Figure \ref{fig:eik22} the admin can see all the stored exams using phpMyAdmin application.

\section{Wordpress plugin Table architecture}

\begin{figure}[h]
\centering
\includegraphics[scale=0.45]{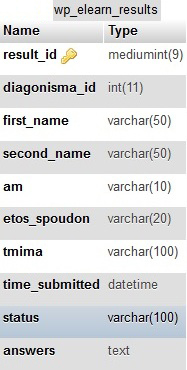} \caption{Plugin's table in which
 the results of student's tests are stored}%
\label{fig:eik26}%
\end{figure}

As it shown in Figure \ref{fig:eik26}, the table of the plugin is named
wp\_elearn\_results. In this table are saved the test results of the students

The following fields are displayed in the table:

Result\_id: The primary key of the result table. (Type: Integer, with a
maximum length of 9 digits)

-Diagonisma\_id: A field in which are stored the identifiers of the tests.
(Type: integer, with a maximum length of 11 digits)

-First\_name: A field in which is stored student's name. (Type: Varchar, with
a maximum length of 50 digits)

-Second\_name: A field in which is stored the last name of the student. (Type:
Varchar, with a maximum length of 50 digits)

-Am: A field in which is stored student's identifier. (Type: Varchar, with a
maximum length of 10 digits)

-Etos\_Spoudon: A field in which is stored the year of the studies of the
student. (Type: Varchar, with a maximum length of 20 digits)

-Tmima: A field in which is stored the department of studies of the student.
(Type: Varchar, with a maximum length of 100 digits)

-Time\_submitted: A field in which is stored the date and the time of the
completed test. (Type: Date and time with the following form YYYY-MM-DD HH:MM:SS)

-Status: A filed in which is stored test's status which can be Open, Checked
or Finalized. (Type: Varchar, with the maximum length of 100 digits)

-Answers: A field in which are stored the answers of the test. (Type: Text,
with a maximum length of 65535 digits)

\section{Functional and non-functional analysis}

\textbf{Functional :}\\
The system supports three different user groups:
\begin{itemize}
	\item The students have the ability to participate in exams
	\item The Teachers have the ability:
	\begin{itemize}
		\item add/edit/delete/view questions and categories of questions
		\item add/edit/delete/view exams
		\item evaluate exams (Grades)
	 \item 	keep an attendance log for each of their tests
	\item mark a student as a successful participant
	\item edit their profile
	\end{itemize}
\end{itemize} 

 \begin{itemize}
 	\item The Administrator has the ability:
	\begin{itemize}
		\item all the abilities of Teacher user group plus editing the teachers' profile
	\end{itemize}
\end{itemize}
\textbf{Non-Functional:}
\begin{itemize}
	\item There is no need for installing additional software as the
	environment works with web browsers so it is accessible
	from anywhere with any operating system.
	\item security and reliability issues.
\end{itemize}

\section{Conclusions}

In the present paper a plugin application (ODES) is developed for the Wordpress platform that manages the examination and auto-grading for students exams and supports conducting exams, collects the answers, auto mark the submissions, and produce the reports for the test and meet the needs of several levels of education (Primary,Secondary,Tertiary etc.) developed with open source software tools and taking into account their advantages and features. In future editions of the ODES application we plan to add the following features:
\begin{itemize}
	\item Link mail to the users and the ability to send notification emails to a single email address, multiple email addresses or a group of WordPress users.
	\item	Show answers at the end of the test or immediately after selection.
	\item Export/Import results to CSV file.
	\item  add multimedia files (images, videos etc.) and other types of questions as well.
\end{itemize}

\end{document}